\documentclass{aastex}
\usepackage{graphicx}
\usepackage{epstopdf}
\usepackage{bm}
\usepackage{lineno}

\newcommand{\be}{\begin{equation}}
\newcommand{\ee}{\end{equation}}
\newcommand{\nn}{\mbox{} \nonumber \\ \mbox{} }
\newcommand{\ba}{\begin{eqnarray}}
\newcommand{\ea}{\end{eqnarray}}
\newcommand{\om}{\omega}

\renewcommand{\v}{{\bf v}}

\newcommand{\Bf}{{magnetic field}}
\newcommand{\Ef}{{electric field}}
\newcommand{\Bfs}{{magnetic fields}}
\newcommand{\NS}{neutron star}

\newcommand{\mss}{magnetospheres}
\newcommand{\NSs}{{neutron stars}}

\newcommand{\EM}{electromagnetic}

\newcommand\eg{{\it{e.g.}}}

\newcommand\lo{\mathrel{\raise.3ex\hbox{$<$}\mkern-14mu\lower0.6ex\hbox{$\sim$}}}
\newcommand\go{\mathrel{\raise.3ex\hbox{$>$}\mkern-14mu\lower0.6ex\hbox{$\sim$}}}

\begin{document}
\title{Nonlinear optics in  strongly magnetized  pair plasma, with applications to FRBs
}

\author{
Maxim Lyutikov\\
 Department of Physics and Astronomy, Purdue University, 525 Northwestern Avenue, West Lafayette, IN, USA 47907
}

\begin{abstract}
Intense radiation field can modify plasma properties, the corresponding refractive index, and lead to such nonlinear propagation effects as self-focusing.
 We estimate the corresponding effects in  pair plasma,  both  in unmagnetized and  strongly magnetically dominated case. 
 First, in the unmagnetized pair plasma the ponderomotive force does not lead to charge separation, but to density depletion. Second, 
for astrophysically relevant plasmas of  pulsar \mss, (and possible  {\it loci} of Fast Radio Bursts),  where cyclotron frequency $\om_B$  dominates over plasma frequency $\om_p$ and the frequency of the \EM\ wave, $\om_B \gg \om_p,\, \om$,  we show that
(i) there is virtually no nonlinearity due to changing effective mass in the field of the wave;
 (ii) ponderomotive force  is $F_p^{(B)} =- {m_e c^2}/({4 B_0^2}) \nabla E^2$; it is reduced by a factor $(\om/\om_B)^2$ if compared to the unmagnetized case ($B_0$ is the external \Bf\ and $E$ is the \Ef\ of the wave); (iii) for radiation beam propagating along  constant \Bf\  in  pair plasma with density $n_\pm$,  the ponderomotive force leads to appearance of circular currents that  lead to the decrease of  the field within the beam by a factor $\Delta B/B_0  = 2\pi n_\pm m_e c^2 {E^2}/{B_0^4}$.
Applications to the physics of FRBs are discussed; we conclude that for parameters of FRB's the  dominant \Bf\ completely suppresses nonlinear radiation effects.
\end{abstract}

\maketitle 

\section{Introduction}

Neutron stars posses magnetic fields that can approach quantum critical \Bf\ \citep[\eg][]{TD93,tlk,2017ARA&A..55..261K}. In addition, pulsars  produce high intensity coherent emission \citep[giant pulses are especially intense][]{popov06} that may modify the properties of the background plasma. The effects  of the back-reaction of the radiation field on the background plasma are becoming even more important with the recent discoveries related to fast radio bursts  \citep{2007Sci...318..777L,2019A&ARv..27....4P,2019arXiv190605878C}, in particularly identifications of the Repeater  FRB121102
\citep{2019ApJ...876L..23H}, FRB180814 \citep{2019Natur.566..235C}, and lately numerous FRBs detected by CHIME \citep{2019arXiv190803507T,2019arXiv190611305J}. Magnetospheres of \NSs\ are one of the main possible  {\it loci} of the FRBs \cite{2016MNRAS.462..941L,2019arXiv190910409L}.

\cite{2019arXiv190103260L} discusses new limitations on the plasma parameters that FRBs impose if compared with pulsars.   High inferred radiation energy densities at the source   renewed interest  in non-linear radiative  phenomena in plasmas \citep{2019MNRAS.489.5688M,2019arXiv191208150G}.  Both of the above cited works consider non-magnetized/weakly magnetized plasma.  As discussed by \cite{2017ApJ...838L..13L,2019arXiv190103260L}, the properties of   first Repeater FRB121102 requires large \Bf\ at the source. For given observed flux and known distance
the  equipartition \Bf\ energy density at the source evaluates 
\be
B_{eq}=   \sqrt {8 \pi  } {\sqrt{\nu F_\nu } D \over c^{3/2} \tau} = 3  \times 10^8 \tau_{-3} ^{-1} 
\, {\rm G}.
\label{B2}
\ee
The resulting cyclotron frequency $\om _B $ is much larger than the  observational frequency and, mostly likely larger than the local plasma frequency $\om_p$. (The inherent assumption is that  the duration of the bursts $\tau \approx 1$ msec $\equiv \tau_{-3}$  is an indication of the emission size.)
Specifically, as \cite{2019arXiv190103260L} argued,  such high \Bfs\ {\it are needed} to avoid high ``normal'' (non-coherent) radiative losses.

As we discussed in the present paper, the radiation-plasma interaction in the case of FRBs takes place in an unusual, compared to the more well studied laboratory laser plasma, regime. First, the plasma is likely to be composer of electron-positron pairs. That eliminates/modifies many effects that arise due to different masses of charge carries even in the unmagnetized case. For example,  in the electron-ion plasma the ponderomotive force leads to electrostatic charge separation. In unmagnetized pair plasma it leads to density depression,  \S \ref{pondenoB}, while in the highly magnetized plasma it leads to the modification of the  background \Bf, \S\ref{pondenoB})
 Most importantly, astrophysical plasma are often   magnetically dominated, so the the cyclotron  motion plays the leading role.

Nonlinear plasma effects in this regime remain unexplored. Highlighting these differences is the main goal of the paper. \citep[Current reviews on relativistically strong lasers, \eg,][do not address this specific regime]{2006RvMP...78..309M,2016JETP..122..426B,1986PhR...138....1S}. 

In this Letter we consider non-linear propagation effects in pair plasma, both non-magnetic and  magnetically dominated plasma with $\om_B \gg \om_p, \om$.

\section{Non-linear effects  in pair plasmas}

There are two main types of non-linear effects that we will consider. First, strong EM wave can induce large relativistic velocities of the plasma particles; this modifies the effective mass and thus changes the dispersion relation. Second, transverse (with respect tot he direction of wave propagation) gradients of wave intensity produce ponderomotive force that modifies the plasma density (or \Bf! - see \S \ref{pondeB}), also changing the dispersion relations. We do not consider plasma effects due to changing intensity of the wave, that lead to longitudinal ponderomotive effects like wake fields.


\subsection{Non-linear effects  in  absence of \Bf}
\label{pondenoB}

In the absence of external \Bf\ a particle
in strong radiation field experiences oscillations  (quiver) with dimensionless transverse momentum \citep{1964PhRv..135..381R,1975OISNP...1.....A,1973PhFl...16.1522K,1976JPlPh..15..335K,PhysRevLett.33.209,Pukhov_2002}
\be
a_0 \equiv \frac{p_\perp}{m_ e c} = \frac{e E} {m_e c \om}
\ee
where $E$ is the \Ef\ in the wave, $\om$ is the frequency of the wave and other notations are standard. When $a\geq 1$ the transverse oscillating momentum of a particle in a wave becomes relativistic. This corresponds to wave's intensity 
\be
P =  a_0^2 \frac{ c E^2}{4\pi} = a_0^2  \frac{m_e^2 c^3 \om^2}{4 \pi  e^2} = 3 \times 10^{14} {\rm erg s} ^{-1} {\rm cm}^{-2}  a_0^2 \nu_9
\ee
where $\nu_9$ is the frequency in $GHz$.

\subsubsection{Non-linear effects  due to changing mass}

In unmagnetized plasma the non-linear effects of the strong laser light can be  first accommodated into changing effective mass of particles \citep{1975OISNP...1.....A}, so that the refractive index $n$  for a circularly polarized electromagnetic wave becomes
\ba &&
n^2 =1 -\frac{ \om_p^2}{\sqrt{1+a_0^2}}
\nn &&
{ \om_p^2}=  \frac{ 4\pi  n_\pm e^2}{m_e}
\ea
where $n_\pm$ is plasma density.
\footnote{Note  that in an electron-ion  plasma with a density $n_\pm$, a displacement of the electrons with respect to the ions generates the electric field $
{E_{disp}} \approx (\om_p/\om)^2{E}
$.
 This does not happen in pair plasma.} 

Consider a beam of radiation propagating in plasma.
 The jump of  the refractive index between the core of the beam and the background due to the changing mass is then
\be
\Delta n \approx  \frac{1}{2}   \frac{ {\om_p}^2}{\om^2} \left( 1- \frac{1}{\sqrt{1+a_0^2}}  \right) \approx
\left\{
\begin{array}{cc}
\frac{1}{4}  \frac{ {\om_p}^2}{\om^2} a_0^2, & a_0\ll 1
\\
\frac{1}{2}  \frac{ {\om_p}^2}{\om^2}  , &  a _0 \gg 1
\end{array}
\right.
\ee
Refractive index is  larger in the core of the beam.
If the radiation pattern forms a beam with decreasing power away from the central axis (this can occur also due to fluctuations on the beam intensity), parameter $a_0$  as well as $n_\pm$ also decrease away from the center,  so that a converging lens is formed.

For $a_0\ll 1$ we can write the refractive index in the form
\ba &&
n=n_0+n_2 E^2
\nn &&
n_0 = 1 - \frac{\om_p^2}{2 \om^2}\approx 1
\nn &&
n_2 = a_0^2  \frac{\om_p^2}{4 \om^2} =  \pi  \frac{n_\pm e^4}{m_e^3 c^2 \om^4}
\ea

If the beam diameter is $ d$, the beam might be expected to expand by diffraction with an angular
divergence of $\theta \sim  \lambda/(d) $. But higher refractive index inside the beam may lead to internal reflection if the total power  of the beam satisfies
\citep{1968SvPhU..10..609A}
\be
P> P_c = \pi a_0^2 \frac{ E^2 c}{4\pi} = \frac{1.22^2 c }{256 n_2} = 7 \times 10^{-4} \frac{ m_e^3 c^5  \om^2}{e^4 n_\pm}
\label{Pc}
\ee
This is total power for self - focusing in unmagnetized plasma, taking only modification of 
mass; weakly non-linear regime is assumed $a_0 \ll 1$.

The corresponding focal length and lensing angle are
\ba &&
R_f \approx \frac{d}{2} \sqrt{ \frac{n_0}{n_2 E^2}}  \approx \frac{d}{a_0} \frac{\om}{\om_p}
\nn && 
\theta_f  = \frac{d} {R_f} = {2} \sqrt{ \frac{n_2 E^2} {n_0}}  \approx \frac{n_2 E^2}{d} \frac{\om_p}{\om}
\label{Rf}
\ea

In the highly non-linear regime, $a_0\gg 1$, the refractive index inside the beam becomes $\approx 1$, while outside it is still $\approx 1 - \frac{ {\om_p}^2}{2\om^2}$. 
Equating the diffraction angle $\sim 1.22 \kappa_{GJ}/(2 a) $ to the critical angle of internal total reflection gives a condition on the width of the self-collimating  beam
\be
d \leq 7.6  \frac{c \omega}{\om_p^2}  = 0.6 \frac{c m_e \om}{ e^4 n_\pm} 
\ee

\subsubsection{Non-linear effects in unmagnetized  pair plasma  due to ponderomotive force}

In addition to changing mass the plasma particles  experiences ponderomotive force due to the  transverse  gradient of the intensity of the wave.  Separating particle motion into fast oscillations along, \eg,  $x$ direction with coordinate-dependent amplitude, 
\ba &&
\ddot{x} = g(x) \cos \om t
\nn &&
g= \frac{e E}{m_e}
\label{ddotx}
\ea
and averaging over fast oscillation, the slow coordinate  $x_0$ evolves according to 
\ba &&
\ddot{x} _0 = - \frac{1}{\om^2} \frac{d}{dx} (g(x)^2) =
 - \frac{e^2}{4 m_e^2 \om^2} \nabla E^2
 \nn && 
{\bf F}_p =-  \frac{e^2}{4 m_e \om^2} \nabla E^2
 \label{ponde}
 \ea
 Relations (\ref{ponde}) give the  drift motion of a charged particle under the effect of a non-uniform oscillating field and the corresponding ponderomotive force $F_p$. 
 
 Typically, in electron-ion  laboratory plasmas the ponderomotive  force on electrons is balanced by electrostatic forces, giving rise to the ponderomotive electrostatic potential
 \be
 \Phi = \frac{m_e}{4 \om^2} g^2 = \frac{e^2 E^2}{4 \om^2}
 \ee
 As the electric fields are screened on Debye/skin depth, this gives the relativistic critical power $ W_c$
 \be 
 W_c \sim P (a=1) \left( \frac{c}{\om_p}\right)^2 = \frac{m_e^2 c^5 \om^2}{  e^2 \om_p^2} 
 \ee
  
  In unmagnetized  pair plasma the situation is very different:  the ponderomotive force (\ref{ponde}) acts both on electrons and positrons, creating a density depression within the beam. This will be balanced by pressure gradients.  In plasma of temperature $T$ the relative density depression within the beam is then
  \be
  \frac{\delta \rho}{\rho} =  \frac{ e^2 E^2}{ m_e T \om^2} = a_0^2 \frac{m_e c^2}{T} =  \frac{a_0^2}{\theta_T}
  \label{rhoo}
  \ee
  where $ \theta_T = {T}/{m_e c^2}$. 
    Thus we expect that in a sufficiently cold  pair plasma, with $\theta_T \leq 1$, a  strong radiation beam creates a density cavity for $ a_0 \sim \theta_T^{1/2}$.

  The density depression (\ref{rhoo})  will create a variation of the refractive index  of the order 
  \be
\Delta n^{(p)} \approx \frac{1}{2}  \frac{ {\om_p}^2}{\om^2} \frac{\delta \rho}{\rho}  \approx    \frac{a_0^2}{\theta_T} \frac{ {\om_p}^2}{\om^2}
\label{dnpond}
\ee
where superscript ${(p)}$ stresses this relation apply to the effects produced by the ponderomotive force.
 
The  refractive index due to the ponderomotive force  is higher in the core of the beam:  effects of increasing effective mass  and density depletion due to the ponderomotive force amplify each other. 
Thus, high intensity radiation field in  unmagnetized pair plasma creates nonlinear lens that focus the light rays,  and lead to further amplification of the energy density of radiation.

Using Eq. (\ref{Pc}) for the critical power in terms of $n_2$,  the expression (\ref{Rf})  for focal length, and 
 Eq. (\ref{dnpond}) for the change of the refraction index, we find in this case
\ba &&
n_2^{(p)}= \frac{\Delta n}{E^2} = \pi \frac{ n e^4}{m_e ^2 \om^4 T}
\nn && 
P_c^{(p)} =7 \times 10^{-2} \frac{ m_e ^3  c^5}{e^4} \frac{  \om^2}{n}  \theta_T = 3  \times 10^8  {\rm erg s}^{-1} \nu_9^2 \lambda_6 ^{-1} b_q ^{-1} P \theta_T
\nn &&
R_f ^{(p)}= \frac{  m_e^{1/2} \om  }{2 \pi^{1/2}  e   n_\pm^{1/2}}  d  \frac{\theta_T^{1/2}}{a_0}
\nn &&
\theta_t ^{(p)}= \frac{2 \pi^{1/2}  e   n_\pm^{1/2}} {  m_e^{1/2} \om  }  \frac{a_0}{\theta_T^{1/2}}
 \ea

 \section{Nonlinearity in magnetically-dominant plasma}
 
  \subsection{No nonlinear effects due to quiver momentum }
\label{quiverB}
 If there is external \Bf\ $B_0$, such that $\om_B \gg \om$, the plasma dynamics changes dramatically. Most importantly,  the leading nonlinear effects in the unmagnetized plasmas - induced by the variation of effective mass - disappears. 
 
 For $\om_B \gg \om$
   a particle in a wave  experiences linear 
 acceleration not for a fraction of wave period, but for a fraction of the cyclotron gyration.
 The magnetic nonlinearity parameter is then
 \be 
 a_0^{(B)} \equiv \frac{p_\perp}{m_ e c} = \frac{e E} {m_e c \om_B} = a_0 \frac{\om}{\om_B} = \frac{ E}{B_0}, 
 \ee
 the ratio of the \Ef\  in the wave to the external \Bf. 
 
 For a wave with energy flux $P$, the  ratio  of the \Ef\ in the wave to the external field is
 \be
  \frac{ E}{B_0}= 2 \sqrt{\pi} \frac{\sqrt{P}}{\sqrt{c} B_0}
  \ee
  It becomes unity for 
  \be
  P= \frac{B_0^2 c}{4\pi} = 4 \times 10^{36}  b_q^2 \mbox{erg cm$^{-2}$ s$^{-1}$}
  \ee
where  we  normalized the \Bf\ to the quantum critical \Bf,
 $B_0 = b_q B_q$, $B_Q= m_e^2 c^3 /(e \hbar)$.  This is unrealistically high energy flux, not likely to be reached:   
the  \Ef\  in the wave is  much smaller than external \Bf: $a_0^{(B)} \ll 1$ \citep[this corrects a typo in][Eq. (5)]{2017ApJ...838L..13L}.

Thus,  instead of large amplitude oscillations a particle experiences an $E \times B$ drift with non-relativistic velocity
\be
\frac{v_\perp}{c}=  a_0^{(B)}  =  \frac{ E}{B_0} \ll 1
\ee
The magnetic non-linearity is always small,  $a_0^{(B)} \ll 1$,  quiver velocity is non-relativistic, and the mass modification in the regime $\om_B \gg \om$ is negligible. 
 
  \subsection{Ponderomotive force across \Bf\ in magnetically-dominant plasma}
  \label{pondeB}
  In pulsars, and presumably FRBs, emission is likely to be produced by relativistic particles propagating   approximately along the local \Bf\ \citep{1969ApL.....3..225R,Sturrock71,1999ApJ...512..804L,Melrose00Review}. 
   Let's assume that the circularly polarized  radiation propagates exactly along  the external \Bf. Typically, in pulsar \mss\ the cyclotron frequency is much higher than the plasma frequency and the radiation frequency (in the plasma frame in the case of relativistic bulk motion). 
   
   As we demonstrated in \S \ref{quiverB},  in the case $\om_ B  \gg \om$ instead of  large amplitude oscillations with $p_ \perp \sim a m_e c$ particles experience ExB drift with velocity $(  E/B_0) c$, where $B_0$ is the external \Bf. 
     Relations for the ponderomotive force  (\ref{ddotx})  and (\ref{ponde})  are then modified
 \ba &&
\ddot{x}^{(B)} = \om  E/B_0 c
\nn &&
g^{(B)}  = \om  \frac{ E}{B_0} c
\nn && 
 {\bf F} _p^{(B)} =- \frac{m_e c^2}{4 B_0^2} \nabla E^2
 \label{ponde1}
 \ea
 where superscript ${(B)}$ indicates that estimate is for the case of strong \Bf. The expression for $F_p^{(B)}$ is the ponderomotive force in the magnetically dominant plasma. 
 
 The ratio of the ponderomotive forces in magnetically dominated plasma  and plasma without \Bf\ is 
 \be
 \frac{F_p^B}{F_p} = \left ( \frac{\om}{\om_B} \right)^2 \ll 1
 \ee
 Thus, the  ponderomotive force is reduced by a factor $(\om/\om_B)^2$ if compared to the unmagnetized case.
 
\begin{figure}[t]
  \centering
  \includegraphics[width=0.49\textwidth]{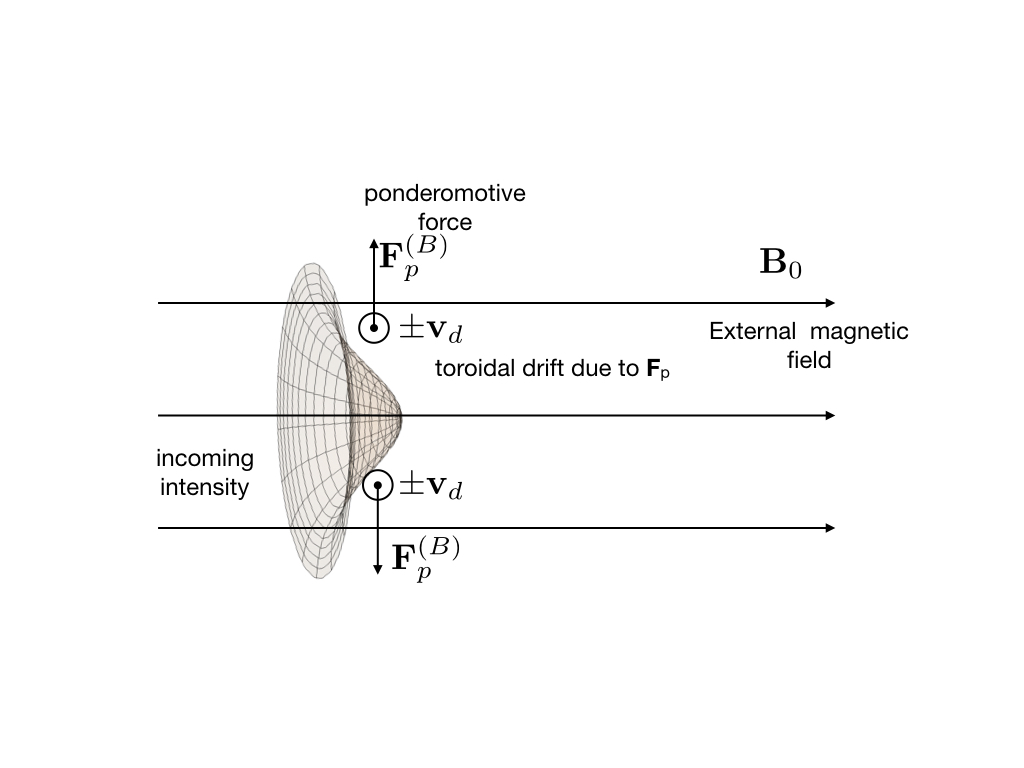}
  \includegraphics[width=0.49\textwidth]{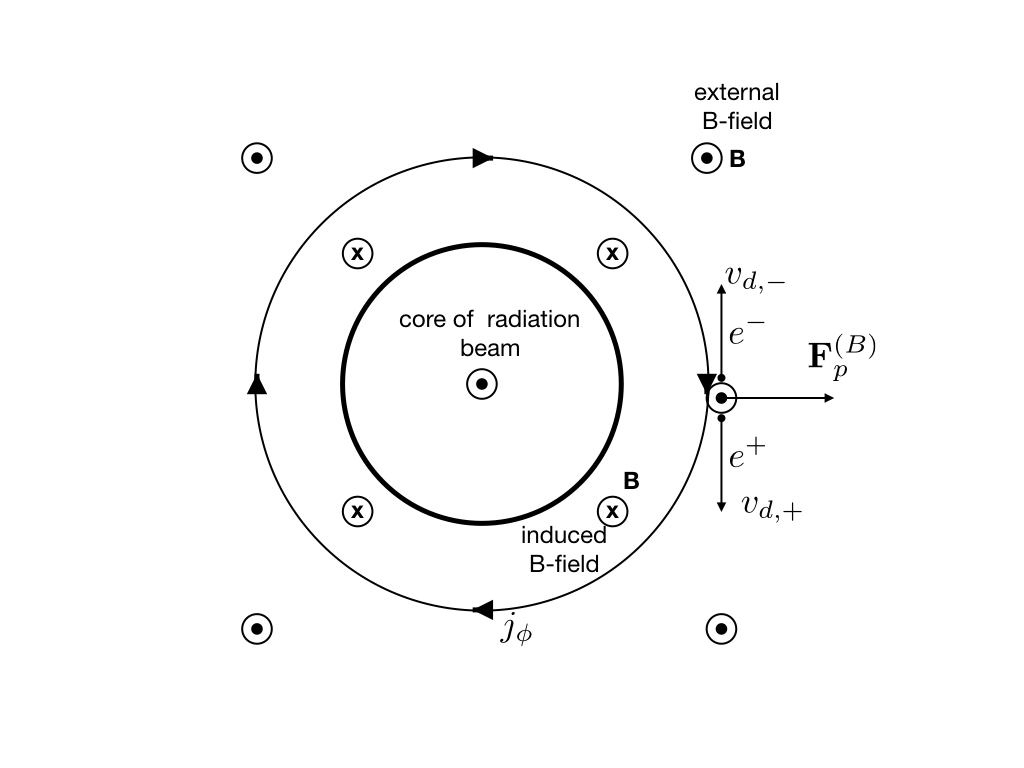}
  \caption{ Left Panel: view from the side. Intense radiation beam is propagating along \Bf. Gradients of field intensity induce ponderomotive force ${\bf F} _p^{(B)}$. In high external \Bf\ $B_0$ the ponderomotive force leads to azimuthal drift of charged particles  $\pm \v_d$ that creates  toroidal current and decrease the background \Bf.  Right panel: view along the direction of the beam. In the core the radiation energy density is high, it induces a ponderomotive force directed away from the center. In the external \Bf\ (chosen to be out of the plane) the ponderomotive force  leads to charge-dependent drift of particles, generation of the toroidal current (in the clockwise direction). The induced current produces a field counter-aligned with the external field}
  \label{pic1}
 \end{figure}

 Most importantly, the effects of the ponderomotive force on the background particles is qualitatively different in the magnetically dominated case as we demonstrate next. 
 If the radiation beam is propagating along \Bf\ and its intensity varies in a perpendicular direction, Eq. (\ref{ponde1}) gives a force on a particle in a direction perpendicular to \Bf.   As a result, the particle will experience a drift with velocity
 \be
 u_{d} = \frac{c}{e} \frac{ {\bf F} _p^{(B)} \times {\bf B}_0}{B_0^2}
 \ee
 The drift is in the azimuthal direction (with respect to the background \Bf),
 see Fig. \ref{pic1}

  Charges of opposite sign rotate in the opposite direction.
 In a  charge-neutral pair plasma with densities $n_\pm$ that will induce a current 
 \be
 j_\phi=  2 e v_d n_\pm
 \ee
The \Bf\ within the beam will be modified  by 
 \be
 \Delta B  = -  2\pi n_\pm m_e c^2 \frac{E^2}{B_0^3}  
 \label{DeltaB}
 \ee
 (\Bf\ is smaller in the core.) 
  
 We can then introduce a magnetic  non-linear intensity parameter $\eta_0^{(B)}$:
 \be
 \eta_0^{(B)} =  \frac{\Delta B}{B_0} =  2\pi n_\pm m_e c^2 \frac{E^2}{B_0^4} = \frac{\om_p^2}{2 \om_B^2} a_0^{(B),2 }
 \ee
Modification of the field becomes of the order of unity at radiative flux 
\be
P ^{(B)} = \eta_0^{(B)}  \frac{ B_0^4}{8 \pi^2 m_e c n_\pm}
\label{PB}
\ee
 dimension of $P ^{(B)} $ is erg cm$^{-2}$ s$^{-1}$.
 
 Modification of the \Bf\ (\ref{DeltaB}) will lead to the changes of the refractive index within a beam (as we argued above  there is no contribution from changing  oscillatory motion of bulk charges).
 In the linear approximation, in the limit $\om_B \gg \om_p, \, \om$,  the wave dispersion reads \citep{AronsBarnard86,Kaz91,1998MNRAS.293..447L,1999JPlPh..62...65L} 
  \be
 n^{(B),2} =1 + \left( \frac{\om_p}{\om_B} \right) ^2
 \ee
 (for parallel propagation; for simplicity we assume cold plasma in its rest frame.)
Expanding in the wave intensity, 
  we find
  \ba &&
  n^{(B)} \approx 1 +  \frac{1}{2} \left( \frac{\om_p}{\om_B} \right) ^2+   \frac{ \om^2 \om_p^4}{2 \om_B^6} a_0^2 =
   \nn &&
  1 +  \frac{1}{2} \left( \frac{\om_p}{\om_B} \right) ^2 +  \frac{ \om_p^4}{2 \om_B^4} a_0^{(B),2} =
  \nn &&
   1 +  \frac{1}{2} \left( \frac{\om_p}{\om_B} \right) ^2+  \frac{ e^2}{2 m_e^2 c^2} \frac{\om_p^4}{\om_B^6} E^2
  \ea
   where $\om_B$ is defined with the  initial background field. The plasma lens  has larger refractive index in the core and thus is  convergent. (The decease in the \Bf\ is due to newly generated internal currents, not expansion, hence the density remains constant.)
   
   Critical power for self-collimation is then
   \be
   P_c^{(B)} = 3 \times 10^{-3} \frac{B_0^6}{m_e ^2 c n_\pm \om^2},
   \label{PB}
   \ee
   and the focal distance and lensing angle
\ba &&
R_f ^{(B)} = \frac{\om_B^3 } {\om \om_p^2} \frac{d}{\sqrt{2} a _0} = \frac{\om_B^2 } { \om_p^2} \frac{d}{\sqrt{2} a _0 ^{(B)} }
\nn &&
\theta_f ^{(B)} ={\sqrt{2} a _0}   \frac{\om \om_p^2}  {\om_B^3 }  =  {\sqrt{2} a _0 ^{(B)} } \frac{ \om_p^2}{\om_B^2 }
\ea

 \subsection{Implications for FRBs}

 Let us use     the properties  of  first Repeater for the estimates of the relevant  parameters \citep{2017ApJ...838L..13L}:
 flux $F_\nu \approx 1$ Jy, frequency $\nu=1$ GHz, distance to the FRB $d_{FRB}\approx $ Gpc, duration $\tau = 1$msec.  The \Ef\ of the wave at the source of size $c \tau$ and the beam power are then
 \ba && 
 E= 2 \sqrt{\pi} \frac{ d_{FRB} \sqrt{ \nu F_nu}} {c^{3/2} \tau} =  2 \times 10^8\, {\rm (in\,  cgs\,  units)}
 \nn &&
 P= \frac{\nu F_\nu d_{FRB} ^2} {c^{3} \tau^2} = 10^{26} \, {\rm erg\,  s}^{-1} \, {\rm cm}^{-2}
 \ea
 \cite[the estimate of the \Ef\  is also the value of the equipartition \Bf,][]{2017ApJ...838L..13L} and Eq. (\ref{B2}).
 The non-linearity parameters then evaluate to
 \ba &&
 a_0^{(B)} = 2 \sqrt{\pi} \frac{ d_{FRB} \sqrt{ \nu F_\nu} }{c^{3/2} \tau B_0} = 4 \times 10^{-6} b_q^{-1}
 \nn &&
 a_0= 5 \times 10^5
 \ea
 Thus, the nonlinear effects are suppressed by the \Bf\ by some ten  orders of magnitude (for quantum field  $b_q=1$).
 
To proceed further we need to estimate  the plasma density. As the sources of  FRBs remain mysterious, below we scale density according to two somewhat oppositely extreme limits: (i)  to the \cite{GoldreichJulian}  density (with some multiplicity $\kappa_{GJ}$); (ii) the quantum  
 density of inverse Compton length cubed,  $n_\pm= \kappa_C \lambda_C^{-3}$, $\lambda_C = \hbar/(m_e c)$.
  These two limits exemplify the clean/light \mss\ of pulsars, and  heavy pair-loaded \mss\ one expects in magnetar flares. 
  
  {\bf Pulsar-like scaling}.
   Using 
   \cite{GoldreichJulian} scaling for plasma density, 
\be
n_\pm = \kappa_{GJ}\frac{\Omega B}{2\pi e c}
\ee
where $\kappa_{GJ}$ is plasma multiplicity and $\Omega$ is the spin frequency of a pulsar, we find

\ba && 
  P_c=  3 \times 10^3 \kappa _{GJ, 6}^{-1} b_q ^{-1} P_{-3}^{1} \, {\rm erg\,  s}^{-1} \, {\rm cm}^{-2}
  \nn &&
  P_c^{(p)} =3 \times 10^5  \kappa _{GJ, 6}^{-1}  \theta_T b_q ^{-1}  P_{-3}^{1} \, {\rm erg\,  s}^{-1} \, {\rm cm}^{-2}
  \nn &&
   P_c^{(B)} =2 \times 10^{60} b_q ^4 P_{-3}^{2} \kappa _{GJ, 6}^{-2} \, {\rm erg\,  s}^{-1} \, {\rm cm}^{-2}
   \nn &&
\theta_f ^{(B)}=  10^{-16} b_q ^{-2} P_{-3}^{-1} \kappa _{GJ, 6}  
   \ea

  {\bf Magnetar-like scaling}.   
  Scaling $ n = \kappa_{C} \lambda_C^{-3}$, we find, using and (\ref{Pc}), (\ref{PB}) 
  \ba && 
  P_c= 5.9 \times 10^{-7}  \kappa _{C}^{-1} \, {\rm erg\,  s}^{-1} \, {\rm cm}^{-2}
  \nn &&
  P_c^{(p)} =6 \times  10^{-5} \kappa _{C}^{-1} \theta_T \, {\rm erg\,  s}^{-1} \, {\rm cm}^{-2}
  \nn &&
   P_c^{(B)} =7 \times 10^{40} b_q ^6 \kappa _{C}^{-2} \, {\rm erg\,  s}^{-1} \, {\rm cm}^{-2}
   \nn &&
   \theta_f ^{(B)}= 5 \times  10^{-7} b_q ^{-3} \kappa _{C}  
   \ea
   where the \Bf\ was scaled to the critical quantum field.
  
  The above estimates cover a wide range of densities and \Bfs.  Yet there is a clear conclusion: the nonlinear effects are highly suppressed in the magnetically-dominant plasma, by some fifty orders of magnitude both for magnetar-like and pulsar-like scaling.


 \section{Conclusion}
 
 In this paper we give estimates of the nonlinear optical effects in strongly magnetized  pair plasma. Two contributing effects are taken into account: effects of the relativistically strong wave on the effective mass of plasma particles and the ponderomotive effects due to the transverse (with respect to the direction of propagation) gradient of the wave's intensity. 
   In the unmagnetized case, both the relativistic  decrease of the effective particle mass, and the ponderomotive effects   lead to the formation of convergent lenses that tend to focus the radiation, further amplifying the non-linearity.  This can be considered as an instability of the radiation front to filamentation. In pair plasma the ponderomotive force leads to density depletion (as opposed to formation of electrostatic potential).
 
 In   magnetically dominated  plasma, \eg\ in the case of {\NS}'s \mss\ and presumably FRB's {\it loci}, where  $\om_B \gg \om, \, \om_p$,  the dynamics is very different. We find that in this regime: (i)  the relativistic effective mass-changing effects on the  wave nonlinearity is completely negligible;
 (ii) 
 the ponderomotive force is suppressed by a factor $(\om/\om_B)^2 \ll 1$ if compared with unmagnetized regime; (iii)   ponderomotive force induces toroidal currents that modify  (decrease) the background \Bf; the resulting lens  is also converging.  
 
  Overall,   the plasma non-linearity is highly suppressed in the magnetized case. As a result,  effects like self-collimation and plasma filamentation  are not likely to play an important role in pulsar \mss, and FRBs.

\section*{Acknowledgments}
This work had been supported by DoE grant DE-SC0016369 and
NASA grant 80NSSC17K0757. I would like to thank Sasha Philippov and Sergey Bulanov for discussions. 

 
 \bibliographystyle{apj}

   \bibliography{/Users/maxim/Home/Research/BibTex}

\end{document}